\documentclass[doublecol,figures,final]{epl2}
\usepackage{amsfonts,amsmath,amssymb}
\usepackage{calrsfs}
\makeatletter
\newcommand{\partialslash}{\mathpalette\@partialslash\relax}
\newcommand{\@partialslash}[1]{%
  \ooalign{\raise.2ex\hbox{$#1\mkern1mu/$}\cr$#1\partial$\cr}}
\makeatother
\renewcommand\today{\ifcase\month\or
  January\or February\or March\or April\or May\or June\or
  July\or August\or September\or October\or November\or December\fi
  \space\number\day, 2010}
\usepackage[ulem=normalem]{changes}
\setauthormarkup[right]{}
\definechangesauthor{R}{red}
\definechangesauthor{M}{magenta}
\definechangesauthor{B}{blue}
\usepackage{upgreek}
\renewcommand{\sigma}{\upsigma}
\renewcommand{\lambda}{\uplambda}
\DeclareMathAlphabet{\foo}{OT1}{ptm}{m}{it}
\newcommand{\g}{{\foo g}}
\usepackage[colorlinks=true,unicode=true,linkcolor=blue,anchorcolor=blue,runcolor=blue,urlcolor=blue,citecolor=blue,breaklinks=true]{hyperref}
\newcommand{\REM}[1]{}
\begin{document}
\title{Quantum critical scaling and the Gross--Neveu model \\
	in $\bm{2+1}$ dimensions}
\author{H. Chamati\thanks{E--mail: \email{chamati@issp.bas.bg}} \and
	N. S. Tonchev\thanks{E--mail: \email{tonchev@issp.bas.bg}}}
\shortauthor{H. Chamati and N. S. Tonchev}
\institute{Institute of Solid State Physics,
Bulgarian Academy of Sciences, 72 Tzarigradsko Chauss\'ee, 1784 Sofia, Bulgaria
}

\abstract{
The quantum critical behavior of the $2+1$ dimensional
Gross--Neveu model in the vicinity of its zero temperature critical
point is considered. The model is known to be renormalisable in the
large $N$ limit, which offers the possibility to obtain expressions
for various thermodynamic functions in closed form. We have used the
concept of finite--size scaling to extract information about the
leading temperature behavior of the free energy and the mass
term, defined by the fermionic condensate and determined the
crossover lines in the coupling ($\g$) -- temperature ($T$) plane.
These are given by $T\sim|\g-\g_c|$, where $\g_c$ denotes the critical
coupling at zero temperature. According to our analysis no
spontaneous symmetry breaking survives at finite temperature. We have found that the leading
temperature behavior of the fermionic condensate is proportional to
the temperature with the critical amplitude $\frac{\sqrt{5}}3\pi$. The
scaling function of the singular part of the free energy is found to
exhibit a maximum at $\frac{\ln2}{2\pi}$ corresponding to
one of the crossover lines. The critical amplitude of
the singular part of the free energy is given by the universal number
$\frac13\left[\frac1{2\pi}\zeta(3)-\mathrm{Cl}_2\left(\frac{\pi}3\right)\right]=-0.274543...$,
where $\zeta(z)$ and $\mathrm{Cl}_2(z)$ are the Riemann zeta and
Clausen's functions, respectively. Interpreted in terms the
thermodynamic Casimir effect, this result implies an attractive
Casimir ``force''. This study is expected to be useful in shedding
light on a broader class of four fermionic models.
}


\maketitle

\section{Introduction}
Quantum phase transitions take place at zero
temperature by varying some nonthermal parameter, say $\g$, such as
composition or pressure, and are driven by genuine quantum fluctuations.
At rather small (when compared to
characteristic excitations in the system) temperature the
singularities of the thermodynamic quantities are altered. It is then
expected that the leading
$T$ dependence of all thermodynamic observables is specified by the properties
of the zero-temperature critical points, which take
place in quantum systems.
In the close vicinity of a second order quantum phase transition
the coupling of statics and dynamics introduces an effective
dimensionality which depends upon (imaginary) time in addition to space
\cite{hertz1976}.
In this case the inverse temperature acts as a finite size
in the imaginary time direction for the quantum system at its
critical point. This allows the investigation of scaling laws for
quantum systems near the quantum critical point in terms of the theory
of finite--size scaling~\cite{sondhi1997,sachdev1999,brankov2000}.

The so called critical or thermodynamic Casimir effect, announced by
Fisher and de Gennes in their study on critical binary liquids
\cite{fisher1978}, is tightly related to the fluctuations of the order
parameter in confined systems exhibiting temperature driven second
order phase
transitions in the bulk \cite{krech1994,brankov2000,tonchev2007a}. The
named fluctuations give rise to a stress inside the confining walls,
which impose effective boundary conditions on the system, depending on
the behavior of the order parameter at the walls. The resulting stress can be
described by the so called Casimir force, obtained as the derivative
of the free energy with respect to the separation between the bounding walls.
The critical Casimir force is a universal quantity, in the sense that it is
independent of the details of the system. These are screened by
critical fluctuations that are correlated over a long
distance, namely the correlation length, that grows
indefinitely as we approach the critical point.
By now it is well established that the Casimir force is
attractive for symmetric boundary conditions (periodic, Dirichlet,
Neumann), while it is repulsive for non symmetric ones (antiperiodic,
Dirichlet--Neumann) \cite{krech1994,brankov2000,chamati2008c}.

In analogy to the critical Casimir effect due to spatial confinement,
we consider the critical behavior of quantum systems in the context of
the critical Casimir effect, defining
the temporal Casimir effect
\cite{chamati2000c,chamati2009} also dubbed \textit{Casimir effect in
time} \cite{palova2009}, where the confinement is
caused by the boundary conditions along the \textit{finite} imaginary time
direction. The singular part of the free energy of a $d$
dimensional system in the vicinity
of its quantum critical point, with hyperscaling satisfied, is expressed as
\begin{equation}\label{qcas}
f_s(\g,T)\approx T^{d+1}\mathfrak{Y}\left(c\mathfrak{g}T^{-1/\nu}\right).
\end{equation}
where $\mathfrak{g}$ and $\nu$ measure, respectively, the distance from the
quantum critical point $\g_c$ and the divergence of the
correlation length. The number
$c$ is a non universal quantity encoding the irrelevant details of the
quantum system. The function $\mathfrak{Y}(x)$ is a
universal scaling function, whose expression depends upon
the universality class of the quantum critical behavior. At $\g=\g_c$,
the universal critical amplitude
$\mathfrak{Y}(0)$ is the Casimir amplitude.
Its sign indicates whether 
the fluctuation induced ``force'' is attractive or repulsive.
In expression (\ref{qcas}) we assumed that the dynamic critical exponent
$z=1$.
Let us mention that the critical Casimir amplitude in
quantum systems is related to the specific heat and can be obtained
from experimental data \cite{brankov2000,castroneto1993}.

In this letter we tackle the quantum critical scaling behavior in
interacting massless fermions in order to make contact with the
thermodynamic Casimir effect in time in such systems.
To achieve our goal we employ the $O(2N)$
Gross--Neveu model \cite{gross1974} in $2+1$ dimensions. The model
describes discrete chiral
symmetry breakdown. Its Hamiltonian reads
\begin{equation}\label{gnham}
\mathcal{H}=\int d^2\bm{x}\left[
-i\bar{\bm\psi}\left(\gamma^1\frac{\partial}{\partial x_1}
+\gamma^2\frac{\partial}{\partial x_2}\right)\bm\psi
-\frac{\g^2}{2N}(\bar{\bm\psi}\bm\psi)^2
\right],
\end{equation}
where the fields $\bar{\bm\psi}\equiv\bar{\bm\psi}(\bm{x},t)$ and
$\bm\psi\equiv\bm{\psi}(\bm{x},t)$ are massless Dirac fermions with
$\bar{\bm\psi}\bm\psi=\sum_{\alpha,i}^N\bar\psi_\alpha^i\psi_\alpha^i$,
$\gamma^\mu$
are Dirac matrices and $\g$ -- a nonzero coupling between
fermions.

Model (\ref{gnham}) may have applications
that can be encountered in a variety of physical phenomena spanning
condensed matter (superconductivity, cold matter, antiferromagnetism)
and elementary particles.
The model is known to be exactly solvable in the large $N$ limit,
using variational calculations \cite{rosenstein1991,moshe2003}.
It can be useful in shedding light on phenomena one might expect
in more realistic models.

The thermodynamics of (\ref{gnham}) in the large $N$ limit and at
finite temperature has been investigated in many occasions (See references
\cite{rosenstein1991,moshe2003} and references therein).
At zero temperature it exhibits a
spontaneous symmetry breaking at a critical coupling $\g_c$. Further,
investigations at finite temperature in the presence of quantum
effects have shown that, under certain assumptions related
to the variational problem (i.e., if only nonnegative
solutions of the corresponding self--consistency or gap equation have
been taken into account), a thermal driven phase transitions takes place
at $T_c>0$. In the neighborhood of $T_c$, the critical
behavior of a number of thermodynamics quantities has been determined.

The aim of the
present study is to go beyond these works and explore the
mutual influence
of quantum and classical fluctuations in the 2+1
Gross--Neveu model in the large $N$ limit in the vicinity of $\g_c$
keeping the temperature finite ($T\to 0^+$).
To this end we determine the scaling behavior of the correlation length and
the free energy. In our analysis we release the constraint
related to the variational problem mentioned above. We will see that this
leads to a ``bona fide'' solution for the free energy with reasonable
physical output.
To the best of our
knowledge this is the first time ever that such a study of the model
under consideration is performed.

The remainder of this letter is organised
as follows: In the next Section, we
recall the main characteristics of the Gross--Neveu model and derive
its critical properties in the vicinity of the quantum critical point
(zero temperature critical coupling), keeping the temperature finite.
To this end we compute the free energy and the gap equation to the leading order
in the large $N$ limit and analyse their corresponding
behaviors as a function of the temperature. In the following Section we
investigate the critical Casimir effect in time, by deriving the
scaling function of the free energy density and extracting the Casimir
amplitude due to the confinement in the time direction.
We conclude the paper by giving a brief account of our results in
the last Section.

\section{Quantum critical behavior}\label{model}
To gain insights into the thermodynamics of model (\ref{gnham})
it is more convenient to use the field theoretic
formulation with the Lagrangian density
\cite{gross1974,rosenstein1991,moshe2003}
\begin{equation}\label{gnmodel}
\mathcal{L}(\bar{\bm\psi},\bm\psi)=
i\bar{\bm\psi}
\partialslash\bm\psi+\frac{\g^2}{2N}(\bar{\bm\psi}\bm\psi)^2.
\end{equation}

The partition function is computed from the path integral
\begin{equation}\label{zpar}
\mathcal{Z}=\int\mathcal{D}\bm\psi\mathcal{D}\bar{\bm\psi}
\exp{\left[-\int
dtd^2\bm{x}\mathcal{L}(\bar{\bm\psi},\bm\psi)\right]},
\end{equation}
where the fermionic fields are antiperiodic in the imaginary time
direction with period
$\beta=T^{-1}$ i.e. $\bm\psi(\bm{x},\beta)=-\bm\psi(\bm{x},0)$. Here
and below we work in units such as $\hbar=k_B=1$.
The discrete chiral symmetry corresponds to
\begin{equation}
\bm\psi=\gamma_5\bm\psi, \qquad \bar{\bm\psi}=-\bar{\bm\psi}\gamma_5,
\end{equation}
defining the fermionic condensate
$\langle\bar{\bm\psi}\bm\psi\rangle$, with thermodynamic average
taken with the action in eq. (\ref{zpar}), as the order parameter or mass
term of the
theory. Under the above transformation the Gross--Neveu Lagrangian is invariant.

In the large $N$ limit, instead of (\ref{gnmodel}), using a standard
decoupling technique \cite{moshe2003} based on the Hubbard--Stratonovich transformation
we get the equivalent effective Lagrangian:
\begin{equation}\label{gneff}
\mathcal{L}_\mathrm{eff.}=i\bar{\bm\psi}\partialslash\bm\psi
+\lambda\bar{\bm\psi}\bm\psi-\frac{N\lambda^2}{2\g^2}.
\end{equation}
where $\lambda\equiv\lambda(\bm{x},t)$ is an auxiliary scalar field.
Let us note that in ref. \cite{rosenstein1989a} 
it was shown that the model is renormalisable order by order in
$\frac1N$ expansion.

The functional integral (\ref{gneff}) entering the expression of the partition function is
Gaussian in fermionic fields. These are integrated out to yield the
effective action
\begin{equation}\label{seff}
\mathcal{S}_\mathrm{eff.}=\frac N{2\g^2}\int_0^\beta dt\int d^2\bm{x}
\lambda^2-N\mathrm{Tr}\ln\left[\partialslash_\mu+\lambda\right].
\end{equation}

With the aid of eq. (\ref{seff}) we may compute the partition
function
using the saddle point method. Thus we obtain an
expression for the free energy to the leading order in $\frac1N$
expansion as:
\begin{equation}\label{freeene}
\frac1N f(\g,T)=\frac{\sigma^2}{2\g^2}-\frac T2\sum_{n=-\infty}^\infty
\int\frac{d^2\bm{q}}{(2\pi)^2}\ln\left[\omega_n^2+\bm{q}^2+\sigma^2\right],
\end{equation}
where $\omega_n=(2n+1)\pi T$ are the Matsubara frequencies for fermions
and $\sigma\equiv\langle\lambda(\bm x,t)\rangle$
is the solution of the gap equation
\begin{equation}\label{selfc}
\frac\sigma{\g^2}=T \sum_{n=-\infty}^\infty
\int\frac{d^2\bm{q}}{(2\pi)^2}\frac\sigma{\omega_n^2+\bm{q}^2+\sigma^2},
\end{equation}
emanating from equating the derivative, with respect to $\sigma$, of
the free energy $\frac1Nf(\g,T)$ to $0$.
For our further analysis we will regularise the theory by introducing
a cutoff $\Lambda$ over the spatial wave vector $\bm{q}$.

Eq. (\ref{selfc}) has a trivial solution $\sigma=0$ and a
nontrivial one $\sigma\neq0$. The stability of each solution is to be
determined by comparing the corresponding free energies. Thus one can
construct the phase diagram of the model under consideration.

To extract the quantum critical behavior of the model (\ref{gneff})
at zero temperature,
we write (\ref{selfc}) in a more tractable form,
employing the Schwinger representation in conjunction with the identity
\begin{equation}
\sum_{n=-\infty}^\infty e^{-(n+\frac12)^2z}=
\sqrt{\frac\pi z}\sum_{n=-\infty}^\infty\cos(n\pi) e^{-n^2\frac{\pi^2}z}
\end{equation}
that can be derived via the Poisson summation formula
\cite{whittaker1996}. Then we have
\begin{equation}\label{finitet}
\frac1{\g^2}=\frac12\int\frac{d^2\bm{q}}{(2\pi)^2}\frac1{\sqrt{\sigma^2+\bm{q}^2}}
+\frac{\sigma}{4\pi}-\frac T{2\pi}\ln2\cosh\frac{\sigma}{2T},
\end{equation}
where we have omitted the trivial solution $\sigma=0$.

Setting $T=0$ in (\ref{finitet}), we find that the quantum critical
point takes place at the critical coupling $\g_c$, to be determined
from
\begin{equation}\label{critg}
\frac1{\g_c^2}=\frac12\int\frac{d^2\bm{q}}{(2\pi)^2}\frac1{|\bm{q}|}.
\end{equation}
The quantum critical region around $\g_c$ is defined by the condition
$|\g-\g_c|\ll1$.
From the behavior of the local zero temperature fermionic pair
correlation function
$\langle\bar{\bm\psi}(\bm{q})\bm\psi(\bm{-q})\rangle\approx(\sigma^2+\bm{q}^2)^{-1}$
at the quantum critical point, we are able to determine the critical exponent
$\eta=0$ and to identify $\sigma^{-1}$ to the
correlation length $\xi_\psi$ to the leading order in the large $N$
expansion. On the other hand the mass term,
$\mathfrak{m}=\langle\bar{\bm\psi}\bm\psi\rangle$, is given by
\begin{equation}
\mathfrak{m}=4\pi\left(\frac1{\g_c^2}-\frac1{\g^2}\right).
\end{equation}
Hence the critical exponent $\beta=1$ and consequently $\nu=1$ to the
leading order in $\frac1N$. Other
critical exponents can be deduced via the hyperscaling relation.

Now we will turn our attention to the properties of the model in a
close vicinity of the quantum critical point $\g_c$, keeping the
temperature finite. For this purpose in lieu of eq.
(\ref{finitet}), we will study the behavior of the mass term
at finite temperature, say $\mathfrak{m}_T$, using equation
\begin{equation}\label{y2d}
\frac1{\g_c^2}-\frac1{\g^2}=\frac T{2\pi}\ln2\cosh\frac{\sigma}{2T}
\end{equation}
obtained via dimensional regularisation. The solution of this equation
is well known
\begin{equation}\label{sigmat}
\sigma_T=2T\mathrm{arccosh\left(\frac12e^{2\pi\varkappa}\right)},
\end{equation}
where we have introduced the scaling variable
\begin{equation}\label{sv}
\varkappa=\frac1T\left(\frac1{\g^2_c}-\frac1{\g^2}\right).
\end{equation}
Let us note that depending on $\varkappa$, $\sigma_T$ can take real or
complex values.

\begin{figure}[ht!]
\begin{center}
\resizebox{\columnwidth}{!}{\includegraphics{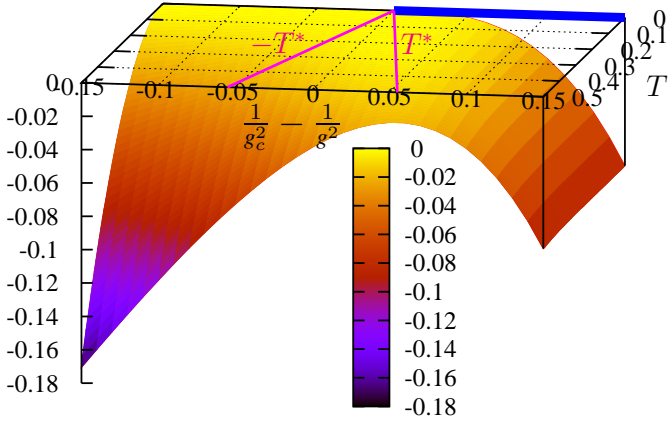}}
\end{center}
\caption{
The free energy as a function of the temperature
and the distance from the quantum critical point. The thick line
corresponds to the zero temperature ordered phase, while $T^*$ shows a
crossover line obtained by setting $\sigma_T=0$ in eq. (\ref{y2d}). The
crossover line $-T^*$ is shown as well.
}
\label{free3d}
\end{figure}

Computing the free
energy with both the trivial solution, $\sigma_T=0$, and the nontrivial one
from eq. (\ref{sigmat}), we find that the phase with the nontrivial solution is
stable at any $T>0$. The free energy corresponding to this
phase is shown in fig. \ref{free3d}. It has a maximum at $T^*$,
which depends on the coupling $\g$, where it coincides with that
evaluated using the triavial solution.
In the strong coupling region, $\g>\g_c$, subtracting eq.
(\ref{finitet}) from its zero temperature counterpart
we obtain
$$
\mathfrak{m}=2T\ln2\cosh\frac{\sigma}{2T}.
$$
By setting $\sigma=0$ we get
\begin{equation}\label{tstar}
T^*=\frac{\mathfrak{m}}{2\ln2},
\end{equation}
wherefrom we conclude that the crossover temperature $T^*$ is proportional to the
zero temperature mass term $\mathfrak{m}$.
We would like to
notice that $T^*$ coincides with the critical
temperature of ref. \cite{rosenstein1989b}, where the solution
$\sigma_T$ in (\ref{sigmat}) for the variational parameter was assumed
to be positive i.e. $\sigma_T>0$.

Now we will explore the temperature dependence of the mass
term in the different regions of the $(\g,T)$ plane, including complex
values for $\sigma_T$ as well.
In the region $\varkappa\to\infty$, by expanding the r.h..s of eq. (\ref{sigmat})
we get
\begin{equation}\label{Tto0}
\sigma_T=\mathfrak{m}-2Te^{-\mathfrak{m}/T}+O\left(e^{-2\mathfrak{m}/T}\right).
\end{equation}
This shows that in this region we have essentially the zero
temperature behavior with exponentially small correction in
temperature. As we approach the quantum critical point a crossover
occurs at $\varkappa^*=\frac{\ln2}{2\pi}$ (the $T^*$ line on fig
\ref{free3d}). For
$\varkappa<\varkappa^*$ the solution $\sigma_T$ is
complex and seems to have no direct physical meaning.
In particular so long as $\g$ approaches its critical value $\g_c$, the
solution $\sigma_T$ creeps towards
\begin{equation}
\sigma_T^2=-\frac49\pi^2 T^2.
\end{equation}
This solution shows that the relation between the fermionic
correlation length and $\sigma_T$ is not straightforward as in the
case of zero temperature. To get a meaningful result we need to use
other means, such as
the asymptotic behavior of the \textit{local} fermionic pair
function. Using standard techniques
\cite{lebellac2004} we end up with
\begin{equation}
\langle\bar{\bm\psi}(\bm{q})\bm\psi(-\bm{q}))\rangle\approx
\frac1{\omega_0^2+\sigma_T^2+\bm{q}^2}.
\end{equation}
This suggests that the thermal mass term (inverse correlation
length) is defined through
\begin{equation}\label{mass}
\mathfrak{m}_T^{2}=\sigma_T^2+\pi^2 T^2,
\end{equation}
i.e. the temperature dependent variational parameter is
shifted by the lowest mode $\omega_0$. This is in
agreement with the fact that fermions have no zero mode at
finite temperature \cite{moshe2003}.
Note that in the limit of zero temperature
$\mathfrak{m}=\lim_{T\to0}\mathfrak{m}_T$ coincides with
$\sigma=\lim_{T\to0}\sigma_T$ as expected.
A similar situation is encountered in the framework
of the spherical
model confined in a film geometry under antiperiodic boundary conditions
\cite{singh1986,chamati2008c}.

The leading behavior of the mass term at finite temperature in the
region $\varkappa\to0^{\pm}$ is
\begin{equation}
\mathfrak{m}_T=\frac{\sqrt{5}}{3}\pi T+\frac{8}{\sqrt{15}}\pi\varkappa T+O(\varkappa^2).
\end{equation}
By setting $\g=\g_c$ we obtain for the critical amplitude of the mass
term at the quantum critical point the universal number $\frac{\sqrt{5}}{3}\pi$.
This shows that the finite temperature mass term vanishes linearly in $T$ as we
approach the quantum critical point from above on the line $\g=\g_c$ in the phase
diagram. Whence the dynamic critical exponent $z$ for the model under
consideration is $z=1$.

\begin{figure}[ht!]
\begin{center}
\resizebox{\columnwidth}{!}{\includegraphics{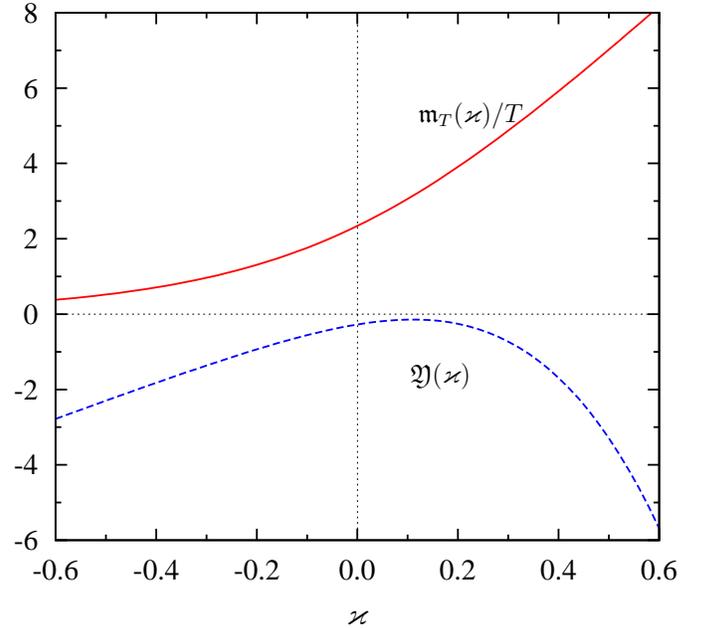}}
\end{center}
\caption{
Dependence of the scaling functions
$\frac{\mathfrak{m}_T(\varkappa)}T$
and $\mathfrak{Y}(\varkappa)$ associated, respectively, to the finite
temperature mass term
(solid line) and the singular part of the free energy
(dashed line) upon the
scaling variable $\varkappa$ in vicinity of the critical
coupling.
}
\label{scaling}
\end{figure}

In the limit $\varkappa\to-\infty$, we have $\sigma_T\approx i\pi T
-iTe^{2\pi\varkappa}$ showing that $\mathfrak{m}_T$ vanishes
exponentially in the weak coupling region, $\g<\g_c$.
The behavior of $\mathfrak{m}_T/T$ as a function of $\varkappa$ is
shown in fig. \ref{scaling}. It is seen that the universal scaling function
associated with $\mathfrak{m}_T$ is a monotonically increasing function
of the scaling variable $\varkappa$.

To sum up,
the mass term (\ref{mass}) normalized to the temperature is found to
behave as:
$$
\frac{\mathfrak{m}_T(\varkappa)}T\approx
\left\{\begin{array}{ll}
\sqrt{2\pi}e^{\pi\varkappa}     & \ \ \varkappa \to -\infty,   \\[5pt]
\frac{\sqrt{5}}3\pi             & \ \ \varkappa = 0,           \\[5pt]
4\pi\varkappa                   & \ \ \varkappa \to \infty,
\end{array}
\right.
$$
as a function of the scaling variable $\varkappa$.
%
Thus we can define three different regions in the $(\g,T)$
plane depending on the behavior of $\mathfrak{m}_T$ as a function of
$\varkappa$. These are:

-- \textit{Renormalized classical}: The mass term diverges exponentially as we
approach the zero temperature phase. For $T>0$, thermal fluctuations
destroy long--range order at any finite temperature.

-- \textit{Quantum critical}: The leading order in $T$ of the mass term is linear.

-- \textit{Quantum disordered}: The mass term is independent of the
temperature.

\begin{figure}[ht!]
\begin{center}
\resizebox{\columnwidth}{!}{\includegraphics{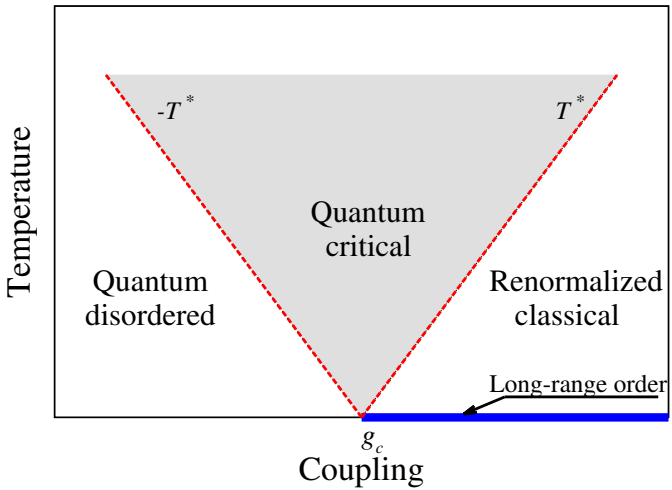}}
\end{center}
\caption{
Qualitative crossover diagram of model (\ref{gnham}) in the quantum critical region.
The thick line shows the zero temperature ordered phase, while the
dashed lines are the crossover lines of fig
\ref{free3d}.
}
\label{diagram}
\end{figure}

These regions are separated by the crossover lines $T^*$ and $-T^*$ on
fig \ref{free3d}. In fig. \ref{diagram} we show the qualitative
crossover diagram of model (\ref{gnham}).
It is worth
noticing that similar behaviors were found in studies involving many other
quantum systems (see e.g. ref. \cite{sachdev1999} and references
therein).

\section{Scaling behavior of the free energy}\label{qcb}
In the quantum critical region, $|\g-\g_c|\ll1$, using the representation
\begin{equation}
\ln z = \int_0^\infty\frac{dt}t\left(e^{-t}-e^{-zt}\right), \qquad z>0,
\end{equation}
and dimensional regularisation, the singular part of the well known
expression for the free energy (\ref{freeene}) can
be written in the scaling form
\begin{equation}\label{fsing}
\frac1Nf_s(\g,T)=T^3 \ \mathfrak{Y}\left(\varkappa \right),
\end{equation}
where we have introduced the universal scaling function
\begin{eqnarray}
\mathfrak{Y}(\varkappa)&=&-\frac12\varkappa\left(\frac{\sigma_T}T\right)^2
+\frac1{12\pi}\left(\frac{\sigma_T}T\right)^3\nonumber\\
& &+\frac{1}{2\pi}\left[\frac{\sigma_T}T \
\mathrm{Li}_2\left(-e^{-\frac{\sigma_T}{T}}\right)+
\mathrm{Li}_3\left(-e^{-\frac{\sigma_T}{T}}\right)\right],\nonumber\\
\end{eqnarray}
in order to make contact with eq. (\ref{qcas}).
Here $\sigma_T\equiv\sigma_T(\varkappa)$ is the solution
(\ref{sigmat}) of the gap equation (\ref{selfc}) and
$$
\mathrm{Li}_\nu(z)=\sum_{k=1}\frac{z^k}{k^\nu}, \qquad |z|<1,
$$
are polylogarithms \cite{lewin1981}.

Let us now explore the behavior of the scaling function
$\mathfrak{Y}(\varkappa)$. Away from the quantum critical point the
leading behavior is given by
$$
\mathfrak{Y}(\varkappa)\approx
\left\{\begin{array}{ll}
\frac1{2}\left[\pi^2\varkappa+\frac1{\pi}\zeta(3)\right]
                                & \ \ \varkappa \to -\infty,   \\[5pt]
-\frac83\pi^2\varkappa^3        & \ \ \varkappa \to \infty.
\end{array}
\right.
$$
This shows that the scaling function is linear in $\varkappa$ in the
limit $\varkappa\to-\infty$ and decreases as $-\varkappa^3$ in the
opposite limit.

In fig. \ref{scaling} we show the behavior of the universal scaling function
$\mathfrak{Y}(\varkappa)$.
It has a maximum estimated to be
$$
\mathfrak{Y}\left(\varkappa^*\right)
=\frac{f_s(\g,T^*)}{\left(T^*\right)^3}=-\frac3{8\pi}\zeta(3)=-0.143485... ,
$$
at $\varkappa=\varkappa^*=\frac{\ln2}{2\pi}=0.110318...$,
corresponding to the crossover line $T^*$ of fig. \ref{free3d}.

At the quantum critical point $\g_c$ we may obtain an analytic
expression using the relations \cite{lewin1981}
\begin{equation}
\mathrm{Li}_3\left(e^{i\frac{\pi}3}\right)=
\frac13\zeta(3)+i\frac{5\pi^3}{162}
\end{equation}
and
\begin{equation}
\mathrm{Li}_2\left(e^{i\theta}\right)=\left[\frac{\pi^2}6-\frac{\theta}4(2\pi-\theta)
+i\mathrm{Cl}_2(\theta)\right]; \ \ \ \  0\leq\theta\leq2\pi,
\end{equation}
where Clausen's function \cite{clausen1832} is defined through
$$
\mathrm{Cl}_2(\theta)=\sum_{n=1}^\infty\frac{\sin(n\theta)}{n^2}.
$$
Thus the critical amplitude of
the singular part of the free energy of the massless Gross--Neveu
model is given by
\begin{eqnarray}\label{casamp}
\mathfrak{Y}(0)&=&
\left[\frac1{6\pi}\zeta(3)-\frac13\mathrm{Cl}_2\left(\frac{\pi}3\right)\right]
\nonumber\\
&=&-0.274543\cdots .
\end{eqnarray}
It is negative.
In terms of the Casimir effect this implies
an attractive Casimir ``force'', despite the fact
that the boundary conditions imposed over the imaginary time are
antiperiodic. A similar attractive behavior for the Casimir force was found in
reference \cite{johnson1975,gundersen1988} induced by the vacuum fluctuations
of fermionic fields between two parallel plates.

\section{Discussion}\label{summary}
We have investigated the quantum critical behavior of the massless
Gross--Neveu model, as a prototype for four--fermionic models.
At
infinite $N$ it has a property that is rarely encountered in the realm
of critical phenomena, namely exact solvability. This property allows
the derivation of thermodynamic functions in closed form, without
recourse to perturbative methods. We have investigated the
critical behavior of the model in the vicinity of the quantum
critical point (zero temperature critical coupling) and
the changes caused by switching on the temperature. In particular we
have paid attention to the interpretation of the quantum
critical scaling in terms of the thermodynamic Casimir effect, resulting from the
confinement in the imaginary time direction.

Notice that the present investigation rules out any phase transition
at finite temperature. Such a transition was obtained in previous
studies assuming that the solution (\ref{sigmat}) of the gap equation obeyed by the variational
parameter of the effective
theory can take only positive real values. Here such a constraint is
released and we consider complex values as well.
This is justified by the fact that the free energy
is always real as shown in figs. \ref{free3d} and \ref{scaling}. In this
case it was found appropriate to redefine the mass term through shifting
the variational parameter by the lowest fermionic mode.

The behavior of the free energy and the mass term in the
vicinity of the quantum critical point was investigated in details. We have
identified the three different regions typical in quantum critical
phenomena: Quantum disordered, quantum
critical and renormalized classical (see fig. \ref{diagram}). These
are separated by crossover lines corresponding to $|\g-\g_c|\sim T$.
In particular, we would like to emphasize that
it was found the mass term is linear in $T$ in
the region $\frac1T|\g-\g_c|\to 0$. The critical amplitude of the mass
term, which coincides with the inverse of the correlation length
describing fermionic correlations, was determined to
be $\frac{\sqrt{5}}3\pi$.

The temperature dependence of the singular
part of the free energy in the vicinity of the quantum critical point
is investigated in details.
Its associated scaling function is found to exhibit a maximum at
$\frac{\ln2}{2\pi}$.
The corresponding critical amplitude may be interpreted as
the Casimir amplitude, due to the confinement in the time
direction. This is
computed exactly [see eq. (\ref{casamp})] and
found to be negative as it is the case for the quantum nonlinear sigma
model \cite{sachdev1993}, although this last model
obeys the Bose--Einstein statistics. This shows that the confinement
in the imaginary time direction is reflected in different way on the sign of the
Casimir amplitude in comparison with the confinement in a space
direction.

\acknowledgments
This work was supported by the Bulgarian Fund for Scientific
Research Grant No. TK-X-1712/2007 (N.T.).

\end{document}